\documentclass{vgtc}                          %

\ifpdf%
  \pdfoutput=1\relax                   %
  \usepackage{graphicx}                %
  \DeclareGraphicsExtensions{.pdf,.png,.jpg,.jpeg} %
\else%
  \usepackage{graphicx}                %
  \DeclareGraphicsExtensions{.eps}     %
\fi%

\graphicspath{{figures/}{pictures/}{images/}{./}} %

\usepackage{microtype}                 %
\PassOptionsToPackage{warn}{textcomp}  %
\usepackage{textcomp}                  %
\usepackage{mathptmx}                   %
\usepackage{times}                     %
\usepackage{cite}                      %
\usepackage{tabu}                      %
\usepackage{booktabs}                  %

\usepackage[final]{changes}

\usepackage{color,soul}

\usepackage{enumitem}
\setlist{nolistsep}
\setlist[itemize]{leftmargin=3mm}
\setlist[enumerate]{leftmargin=5mm}

\usepackage{titlesec}
\titlespacing*{\section}{0pt}{2.5ex plus 1ex minus 5ex}{1.3ex plus .2ex minus 3ex}
\usepackage{color,soul}

\definecolor{outlier}{HTML}{FB8072}
\definecolor{feature}{HTML}{B3DE69}
\definecolor{correlation}{HTML}{8DD3C7}
\definecolor{trend}{HTML}{BEBADA}
\definecolor{topology}{HTML}{80B1D3}

\onlineid{1013}

\vgtccategory{Research}

\vgtcinsertpkg

\title{What-Why Analysis of Expert Interviews: \\ Analysing Geographically-Embedded Flow Data}

\author{Yalong Yang\thanks{e-mail: yalong.yang@monash.edu} %
\and Sarah Goodwin\thanks{e-mail: sarah.goodwin@monash.edu}}%
\affiliation{\scriptsize Caulfield School of Information Technology, Monash University}

\abstract{In this paper, we present our analysis of five expert interviews, each from a different application domain. Such analysis is crucial to understanding the real-world scenarios of analysing geographically-embedded flow data. The results of our analysis show that similar high-level tasks were conducted in different domains. To better describe the targets of these tasks, we proposed three \added{flow-targets} for analysing geographically-embedded flow data: single flow, total flow and regional flow. 
} %

\CCScatlist{
  \CCScatTwelve{Human-centered computing}{Visualization}{Visualization design and evaluation methods}{} 
  \vspace{-0.2em}
}

\begin{document}

\firstsection{Introduction}
\maketitle   
\vspace{-0.2em}
Many domain experts and analysts around the world are interested in discovering insights and patterns related to some form of commodity flow between different geographic locations. Such activities include migration patterns, urban planning, animal movement, diesase mapping, financial trading and commuting behaviour.

Geographically-embedded flow data, often termed spatial movement data, trajectory or origin-destination (OD) data, contains geographical locations (the origin, destination and potentially mid-way points), a connection or trajectory between them and a value referring to the magnitude of the flow between the locations. Data of this type is increasing with advances in technology, yet there are still limited tools to effectively analyse and visualise such data in order to discover patterns and inform decision making. 
In recent years we have seen growing support and solutions for the analysis of flow data, yet to the best of our knowledge, there is no research focusing on identifying real-world tasks conducted by the domain experts.

This work was conducted as part of a 3.5 year PhD project focusing on exploring the design space of geographically-embedded flow visualisation~\cite[Chap. 2]{phdthesis}. The motivation for these interviews was to understand the requirements and role of flow data in real-world applications and existing work flows across different disciplines, in order to help inform new visualisation designs~\cite{Yang:2017cy,Yang:2019bp}.  %

The contributions of this work are: 
1)~three high-level \added{flow-targets} for analysing geographically-embedded flow data;
2)~using Munzner's framework~\cite{Munzner:2014wj} to abstract analytic tasks in interview analysis;
3)~five use cases for motivating flow data visualisation design.

\vspace{-0.2em}
\section{Related Work}
\vspace{-0.2em}
\textbf{Task Taxonomy ---} 
Visualisation tasks have been grouped and classified in a number of examples e.g.~\cite{amar2005low,lee2006task}.
More recently, Brehmer and Munzner compiled these classifications into the \textit{what-why-how} framework~\cite{Brehmer:2013fq,Munzner:2014wj} to systematically abstract work flows for visual analytics. Specific taxonomies for geographical data visualisation has also been explored. Closest to this study is the study of Roth~\cite{Roth:2013ja}, who interviewed expert interactive map users to develop a taxonomy of interaction primitives for map-based visualisation. %
Earlier research by Andrienko~\textit{et al.} also discussed several task taxonomies for analysing spatial-temporal data~\cite{Andrienko:2006ek}. Both Roth and Andrienko~\textit{et al.} distinguished tasks into two levels according to the level of data analysis: \emph{elementary} and \emph{synoptic}. \emph{Elementary tasks} refer to individual elements; \emph{Synoptic tasks}~\cite{Andrienko:2006ek} (refered to as `general' in~\cite{Roth:2013ja}) involve the whole reference set or its subsets. This task taxonomy research inspired our detailed discussion of task targets in Sec.~\ref{sec:interviews:summary}.

\noindent\textbf{Expert Interviews ---} There are several other recent interview studies in visualisation research. Alspaugh~\textit{et al.} interviewed data analysts in a range of industrial, academic, and regulatory environments to understand exploratory activities~\cite{Alspaugh:2018kp}. Kandel~\textit{et al.} interviewed enterprise analysts about their overall analysis process and the organizational context~\cite{Kandel:2012ht}. Both were conducted with a broad focus, while in this paper we concentrate specifically on the analysis of geographically-embedded flow data across different domains. 
\vspace{-0.1em}
\section{Methodology}
\noindent\textbf{Recruitment ---} Potential interviewees were identified and recruited from diverse domains. These experts were chosen because of their knowledge and experience of having analysed 
geographically-embedded flow data during their professional careers (see Tab.~\ref{tab:interview:participants}).
 
\noindent\textbf{Preparation ---} In order to tailor the interview questions to meet the specific needs and expertise of the interviewee and to ensure the interview time was kept to a minimum, prior to each interview each interviewee conducted a preliminary survey. This survey contained three specific sections: \textit{consent}; \textit{personal information}; and \textit{related projects}. We present an overview of these details in Tab.~\ref{tab:interview:participants}.

A semi-structured interview was prepared base on the survey information for each candidate. The basic structure of the interview consisted of the following questions:
\begin{itemize}
	\item Can you give a brief overview of the related project(s)?
	\item Can you explain the data in more detail?
	\item What are/were the initial motivations for the project(s)?
	\item What tasks are involved in the analysis?
	\item Did you use visualisation tools in your analysis? If so, which?
	\item Do you have suggestions on how to improve these tools?
\end{itemize}

\noindent\textbf{Interview Method ---} 
Interviews were conducted via VoIP software, such as Skype or face-to-face when possible. All interviews were audio-recorded and later transcribed. A pilot interview was also carried out internally to test the structure, questions and timing.

\begin{table*}[tb]
    \centering
    \includegraphics[width=0.93\textwidth]{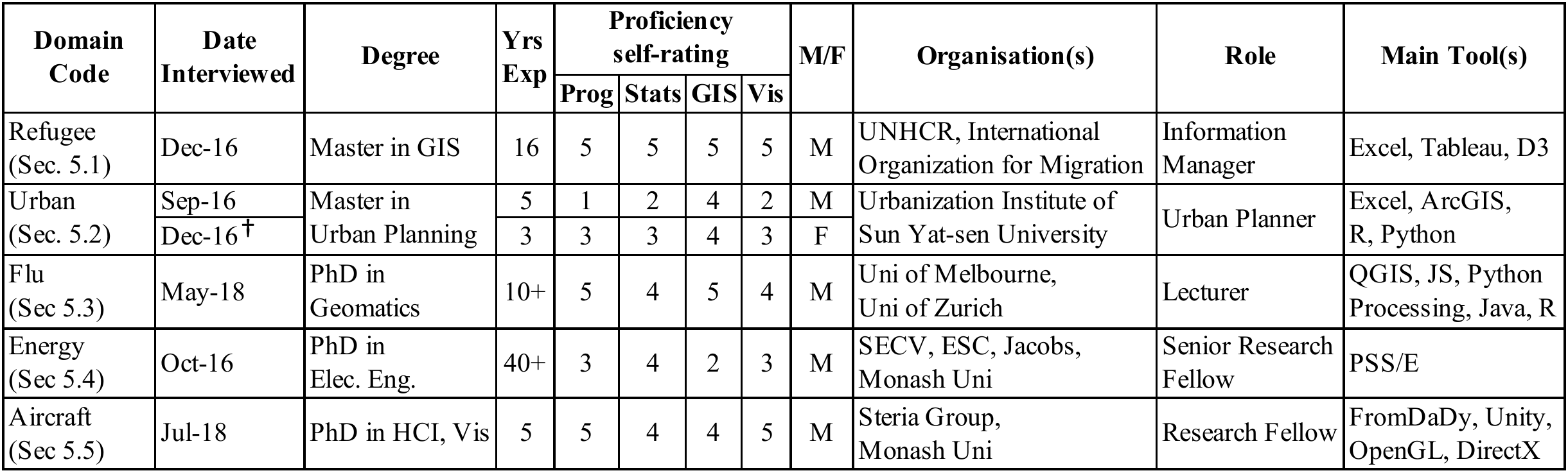}
    \caption{Interviewee background and expertise. Proficiency self-rating range from 1~(No Experience), 2~(Beginner), 3~(Knowledgeable), 4~(Skilled) and 5~(Expert) with clear definition in the survey (see supplementary materials). Abbreviations: UNHCR~(United Nations High Commissioner for Refugees), SECV~(State Electricity Commission of Victoria) and ESC~(Essential Services Commission). ${\dagger}$~Date of presentation (see Sec.~\ref{sec:interviews:urban}).}
    \label{tab:interview:participants}
    \vspace{-2em}  
\end{table*}

\noindent\textbf{Analysis ---} 
\label{sec:interviews:method}
As each project was unique the transcriptions contained very detailed descriptions of each project and whilst the data described in each interview is similar in structure (containing geographical locations and flows between them), the data is analysed and visualised for very different reasons. An initial qualitative coding exercise, similar to~\cite{Alspaugh:2018kp,Roth:2013ja}, revealed the difficulty of using this method of analysis due to diversity of the projects. As an alternative, we chose to analyse the transcripts using the \emph{what-why-how} framework by Munzner~\cite{Munzner:2014wj} as this framework provides us with a structure to think systematically about the similarities in analysis tasks and design choices, within the large and complex design space. %
For this analysis, we paid particular focus on:  
\begin{itemize} 
	\item \textbf{What: Data Abstraction}. Datasets are classified into: tables, networks, fields, geometry (or spatial)~\cite[Chap. 2]{Munzner:2014wj}.
	\item \textbf{Why: Task Abstraction}. Tasks are analysed with: \emph{actions} and \emph{targets}~\cite[Chap. 3]{Munzner:2014wj}. 
	\begin{itemize}
	 	\item There are three levels of \emph{actions}: the highest refers to \emph{consume} or \emph{produce} information; the middle to \emph{finding} the required information; and the bottom to query \emph{targets} at different scopes.
	 	\item  \emph{Targets} refer to the aspect of the data that is of interest to the viewer. There are four kinds of high-level targets: all data, attributes, network data and spatial data.
	 \end{itemize}   
\end{itemize}
A visualisation task can be described as \emph{actions} with a \emph{target}.  
Compared to other analysis methods, this framework guides the translation of domain-specific problems into abstract visualisation tasks using a multi-level typology of tasks. This typology provides abstract and flexible descriptions of tasks which allows useful comparison to be made between different application domains~\cite{Brehmer:2013fq}. Due to very limited visualisation examples provided by the interviewees, we could not analyse the ``How'' part of the framework.

\noindent\textbf{Limitations ---}
The topics and tasks described by the interviewees might be different from those undertaken by other analysts in a similar position in the same domain.  Due to our desire to gain a breadth in understanding, we had to spend time during each interview gaining knowledge of the domain. As each interview was limited to one hour we focused on understanding the high-level tasks, domain and data characteristics. This did not allow time to explore more detailed aspects of the data such as spatial scale, spatial resolution, amount, quality, uncertainty or time dimension. 
Therefore there is a possibility that we have missed tasks through our method of query and analysis. Further interviews and continued discussions could reveal further details and important tasks in each domain.  
\section{Overview}
We briefly introduce the abstracted data and general tasks conducted in the five projects referenced in our interviews. 

\begin{figure}[t!]
    \vspace{-0.1em}
    \centering
    \includegraphics[width=0.8\columnwidth]{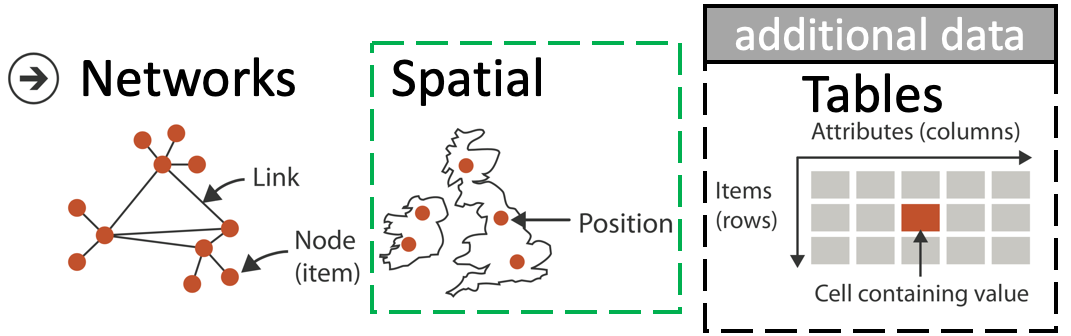}
    \vspace{-0.8em}
    \caption{\emph{What --- data abstraction} for different interviews. Icons derived from~\cite[Chap. 2]{Munzner:2014wj} under CC BY-4.0 Licence.}
    \label{fig:interviews:what} 
    \vspace{-2em}
\end{figure}

\noindent\textbf{What: Data Abstraction ---}
Geographically-embedded flow data can be considered as the combination of \emph{spatial} data and \emph{network} data. In all five interviews these two types of data were mentioned as being collected and analysed. Fig.~\ref{fig:interviews:what} provides a graphic depiction of the data abstractions using icons from~\cite[Chap. 2]{Munzner:2014wj}.

In the case of energy analysis  (see  Sec.~\ref{sec:interviews:energy}), %
geographic patterns in the data are secondary in current work flows, as analysts are more concerned about the network perspective of power grids. 
Thus, we highlight this circumstance using a dashed box in Fig.~\ref{fig:interviews:what}. Additional tabular data were sometimes used to add context (this is shown as a grey dashed box in Fig.~\ref{fig:interviews:what}). For example, analysts investigate the correlation between weather and refugee and migrant arrivals (see Sec.~\ref{sec:interviews:un}) and demographic data is used to inform the design of existing or new airways (see Sec.~\ref{sec:interviews:airline}). 
With the exception of Refugee Movement (Sec.~\ref{sec:interviews:un}), all other interviewees mentioned scale hindering their ability to analyse and visualise the data. 

\noindent\textbf{Why: Task Abstraction ---}
A task is described as the combination of actions and a target. 
Firstly, the highest level of action~---~\textit{\textbf{Analyse}}: from the interviews we identify two main goals: analysts \textit{consume} information to get better insights into the data as well as using other analytic tools to \textit{produce} new information. 
\textit{Consume} tasks mentioned in the interviews mainly related to \emph{presenting}. This is possibly because the interviewees mainly talked about their experience of decision-making, planning and forecasting instead of generating new hypotheses.
\textit{Produce} tasks mentioned in the interviews mainly relate to \emph{deriving} new data. %
For the middle level action~---~\textit{\textbf{Search}}: different specific actions were taken based on whether the identity and the location of the target is known or not. For the lowest level action~---~\textit{\textbf{Query}}: \emph{identify}, \emph{summarise} and \emph{compare} are commonly mentioned in the interviews with different targets.

\begin{figure*}[t!]
    \centering
    \includegraphics[width=1\textwidth]{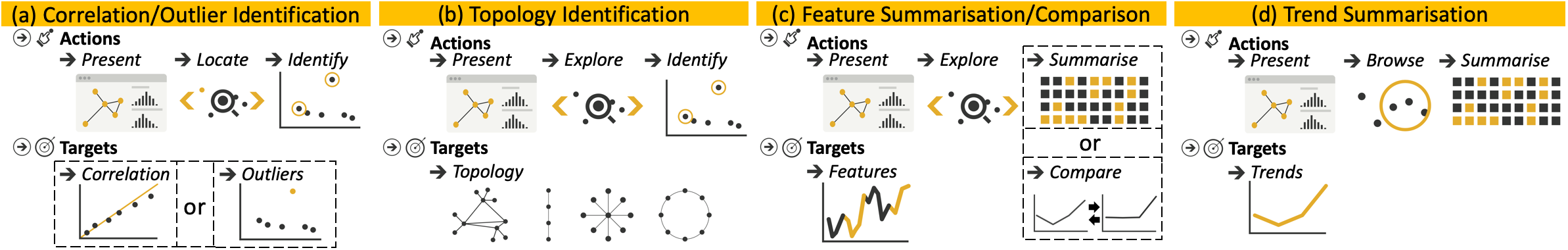}
    \vspace{-1.5em}
    \caption{\emph{Why --- task abstraction} for different interviews. Icons derived from~\cite[Chap. 3]{Munzner:2014wj} under CC BY-4.0 Licence.}
    \label{fig:interviews:why} 
    \vspace{-2em}
\end{figure*}

In terms of common \emph{targets} of interest, the interviewees were interested in: \textit{Trends}, \textit{Outliers}, \textit{Features} and \textit{Topology}. Fig.~\ref{fig:interviews:why} provides a graphic depiction of the task abstraction, using icons from~\cite[Chap. 3]{Munzner:2014wj}. Fig.~\ref{fig:interviews:what} and Fig.~\ref{fig:interviews:why} are referred to throughout the following section, where we describe each individual interview in terms of problem context, data and task abstraction. %

\section{Details}
\label{sec:interviews:details}

\subsection{Refugee and Migrant Movement}
\label{sec:interviews:un}
\vspace{-0.3em}
At times of violent conflict in the world, hundreds of thousands of innocent lives can be affected. In such disastrous situations, many organisations around the world work hard to ensure people's safety. UNHCR and IOM (see Tab.~\ref{tab:interview:participants}) are two of these organisations. They both collect and analyse refugee and migrant data and produce situation reports to support evidence-based decision-making and advocacy on refugee issues.
\deleted{In November 2016, we conducted an interview with Edgar Scrase about his professional experience.}

\noindent\textbf{What: Data Abstraction ---}
Data collection and quality assurance is particularly difficult in this domain. Data collection is a manual process and there is often missing, incomplete or inaccurate data recorded. Collection involves multiple countries, languages and different standards. Data is often not detailed enough and this can cause difficulties during analysis; As stated by the interviewee, ``\textit{UNHCR and IOM conduct regular discussions with governments to attempt to better harmonise this information.}''
 
Important data related to refugee and migrant movement analysis is \emph{arrival data}. This data is collected by local governments or agencies about individuals with the following attributes: \textit{country of origin}; \textit{country of arrival}; \textit{intended destination}; \textit{sex}; \textit{age}; \textit{whether travelling with companion(s), and if so with whom}. In the case of children, whether or not they are unaccompanied or have been separated from their parents. In some cases, refugees and migrants are relocated to another country after their arrival. Such individual records are collected with: \textit{country of arrival} and \textit{of relocation}.

The analysis is usually conducted with aggregated data rather than individual records. There are many different ways to aggregate records e.g. by OD pair, by time period or the combination of them. The analysts are  also interested in estimating future arrivals. One way to achieve short term projections is to investigate the correlation between the refugee and migrant geographical locations and other factors that influence movement, such as the weather.

\noindent\textbf{Why: Task Abstraction ---} 
The main motivation for this work is to more effectively align scarce resources. This aim can be broken into two distinct analytical tasks:
\begin{itemize}
	\item Estimate the arrival. Better estimation can help governments and organisations prepare. There are different scenarios where visualisation can facilitate this task:%
	\begin{itemize}
		\item  Analysts investigate if arrivals in different locations change across time (Fig.~\ref{fig:interviews:why}d: \textbf{\setlength{\fboxsep}{2pt}\colorbox{trend}{Trend} Summarisation});
		\item Analysts investigate possible factors that correlate with arrivals. For example, they found bad weather usually results in fewer arrivals. (Fig.~\ref{fig:interviews:why}a: \textbf{\setlength{\fboxsep}{2pt}\colorbox{correlation}{Correlation} Identification}) 
	\end{itemize} 
 
	\item Prioritise relief operations. Due to limited resources, the most urgent cases, like unaccompanied children, are prioritised. These cases are found through \textbf{\setlength{\fboxsep}{2pt}\colorbox{outlier}{Outlier} Identification} (Fig.~\ref{fig:interviews:why}a).
\end{itemize}
Besides resource allocation, there is also a focus on identifying smuggling. For this, analytical modelling and calculations to \textbf{derive} the data are needed. Then visual analytical tools can help identify possible smuggling routes via \textbf{\setlength{\fboxsep}{2pt}\colorbox{outlier}{Outlier} Identification} (Fig.~\ref{fig:interviews:why}a).

\vspace{-0.5em}
\subsection{Urban Planning: a Bike Sharing Schema}
\label{sec:interviews:urban}
\vspace{-0.1em}
Urban planning focuses on the design of the urban environment, including land, environmental resources and public infrastructures. With the rapidly increased population and size of cities, many challenges emerge and affect people’s daily life. On the other hand, vast amounts of data can be collected from urban environments and activities. Such data can be used to improve urban planning practice.
In this section, we describe a bike sharing schema planning project with the analysis of various datasets. 

\deleted{In September 2016, Yaofu  Huang was interviewed about his} 
The interviewee worked on a bike sharing schema planning project for the city of Xiamen (population of $\sim$3.5 million). The project involved analysing the existing bike sharing schema and planning for an extension to encourage more bike sharing members and ultimately reduce car congestion in the city. Part of the analysis and findings were published in~\cite{zjf2018}. 
In addition, 
\deleted{removed:in December 2016, Jiafen Zheng gave a detailed oral presentation of the project Organised by the Dean of UI-SYSU: Prof. Xun Li} 
the task analysis from the interview in this section was supplemented with additional context from an oral presentation of the project (see Tab.~\ref{tab:interview:participants}).

\noindent\textbf{What: Data Abstraction ---}
\label{sec:interview:urban:what}
One of the main datasets used in the analysis was the bike sharing useage data from the existing schema, in operation since July 2014. This data has two distinct components:
\begin{itemize}
	\item \emph{Docking Stations} consisting of: \emph{geographic location}, \emph{capacity} and \emph{number of docked bikes} in real-time.
	\item \emph{Journey records}, consisting of: \emph{origin docking station}, \emph{start time}, \emph{destination docking station} and \emph{end time}.
\end{itemize}
For context of the city and how the population move around the city, the following datasets were also used:
\begin{itemize}
	\item Road Network: seven different types of roads in Xiamen.
	\item Public Transport: bus stations, routes and travel card journeys.
	\item Taxi Journeys: GPS tracked trajectories.
	\item Region Category: Living, Working and Mixed regions were derived from population and economic census data.
	\item Points of Interests (POIs): locations of important locations such as schools, hospitals, parks, shopping malls, industrial parks etc.
\end{itemize}

\noindent\textbf{Why: Task Abstraction ---}
Trying to encourage people to use bike sharing instead of cars can be achieved by expanding the catchment area (building new docking stations), adding more bikes to the schema (building new and increasing current docking station capacity), and/or facilitating safe cycling in the city by extending existing or constructing new bike lanes.
A key question is how to choose the locations for these stations and lanes to best serve people's travel goals and fit within the constraints of the built environment. 

The previous bike usage data was analysed, and revealed that the majority of previous use was for short distance travel. To explore this further, short distance travel routes in the form of OD and trajectories were extracted from previous usage data, public transport data, and taxi data. To plan for new docking stations, three types of candidates were identified through \textbf{\setlength{\fboxsep}{2pt}\colorbox{outlier}{Outlier} Identification} (Fig.~\ref{fig:interviews:why}a) relating to popular origins or destinations, dense intersections of trajectories and POIs. To plan for new and extended bike lanes, the existing lane network was analysed, and missing short connections between docking stations and other facilities, e.g. POIs or public transit stations (Fig.~\ref{fig:interviews:why}a and c: \textbf{\setlength{\fboxsep}{2pt}\colorbox{outlier}{Outlier} and \setlength{\fboxsep}{2pt}\colorbox{topology}{Topology} Identification}) were identified.  These became candidates for new bike lanes.

In addition to the main aim, the interviewees discussed some other interesting findings\deleted{they discovered in the analysis}, which conform to existing literature on bike sharing schema use in other countries, such as in London, UK~\cite{Beecham:2014he}:
\begin{itemize}
	\item Analysing bike usage and current bike lanes (see Fig.~\ref{fig:interviews:why}a:~\textbf{\setlength{\fboxsep}{2pt}\colorbox{correlation}{Correlation} Identification}), revealed that whilst many of the existing lanes were on main roads, these lanes were hardly used by the bike sharing riders. One possible reason suggested was safety concerns of riding on busy roads.
	\item Through analysing the topology of bicycle usage (Fig.~\ref{fig:interviews:why}c:~\textbf{\setlength{\fboxsep}{2pt}\colorbox{topology}{Topology} Identification}), different patterns of use appear in different types (working, living and mixed) regions.

\end{itemize}

\vspace{-0.5em}
\subsection{Flu Forecasting} 
\label{sec:interviews:indoor}
\vspace{-0.3em}
Flu not only affects people’s daily lives but the local economy and health services. Influenza epidemics vary substantially in size, location, timing and duration from year to year, making it a challenge to deliver timely and proportionate responses~\cite{MOSS:2017il}. Understanding, and predicting epidemics is important to assess the risk and effectively manage the public health resources. 
\deleted{remoevd: Martin Tomko instead} The interviewee works with computational epidemiologists, introducing spatial tracking data into the prediction model aiming at fine-grained spatial resolution, to allow more targeted predictions for resources\footnote{\url{https://networkedsociety.unimelb.edu.au/research/projects/active/urban-flu-forecasting}}. \deleted{removed: Martin was interviewed in May 2018.} At the time of the interview, the project was at an early stage, and the team was focused on exploration. The data described and tasks abstracted describe the data analysis and expected tasks in later stages.

\noindent\textbf{What: Data Abstraction ---}
There are two main datasets in this project: flu notifications and urban mobility. Flu notification data contains weekly (time) updates of positively tested cases of flu by postcode (geographical location). Urban mobility data was acquired from two different sources: a) Australian Bureau of Statistics (ABS) journey to work data; b) Satellite Navigation (satnav) tracking data of around 120,000 users across Australia. All data contains spatial information, while the urban mobility data also form networks. Records are investigated in an aggregated manner, usually by time and/or spatial regions. 

\noindent\textbf{Why: Task Abstraction ---} One task is to investigate ``flu spread'' in a spatial and temporal manner. The time-varying semantic is essential for this task along with the geographical locations of flu reports. By first presenting the change of location of flu reports visually and then connecting these locations heuristically, a network of flu spread structures can be approximated. Subsequently the topology can be analysed based on the dervied geographical networks using the geographical information (Fig.~\ref{fig:interviews:why}c:~\textbf{\setlength{\fboxsep}{2pt}\colorbox{topology}{Topology} Identification}). 
To introduce spatial flow data, they first investigated the spatial correlation between urban mobility and the flu spread, which involves visually comparing two geographic networks and identifying correlation between them (Fig.~\ref{fig:interviews:why}a:~\textbf{\setlength{\fboxsep}{2pt}\colorbox{correlation}{Correlation} Identification}). They now plan to integrate urban mobility mathematically into the existing statistical prediction model.

The interviewee noted that only limited visualisations have been created for this project, and whilst they store the OD data as a matrix they have presented most of their findings as bar and line charts to show changes over time.  Scalability was identified as a potential issue when visualising the urban mobility data, as visual clutter can make it difficult to read useful information.
\vspace{-0.5em}
\subsection{Energy Network Flows}
\label{sec:interviews:energy}
\vspace{-0.3em}
Put simply, a power system has a network of transmission lines connecting power stations and customers. At present electricity is not economically storable. To optimise a power system, the demand from customers needs to be met at every moment~\cite{Bouts:2015fr}. Maintaining the system effectively is becoming even more challenging with embedded micro generation, such as solar panels on customer sites. \deleted{Removed: In October 2016, Ross Gawler was interviewed about his experience in - amended to:} In this interview, we focused on the area of power transmission system analysis, network design and planning.

\noindent\textbf{What: Data Abstraction ---}
There are two high-level components in a power system: \emph{elements} and \emph{transmission lines}. \emph{Elements} is a collective name for transformers, voltage regulators, generators, etc. Different information is captured for different types of devices. Voltage and temperature are commonly measured in real time and the geographic location is recorded. \emph{Transmission lines} are the links connecting \emph{elements}. Along the \emph{transmission lines}, there are two types of flows: \emph{active} and \emph{reactive} power flows. Current is also measured. For emergency alert and analysis, additional datasets are used, e.g weather, to assist the reasoning process.

\noindent\textbf{Why: Task Abstraction ---}
Two main problems for energy flow analysis were described during the interview: a) the operating problem, that is instantaneously adjusting elements to increase the capacity of a portion of the power system; b) the planning problem, which focuses on how to economically relieve constraints where they may occur in the future. 

For both problems, different types of constraints were modelled first based on the \emph{elements} and \emph{transmission lines}:
	a) Thermal: Current in the transmission \deleted{exceeding an element rating} can generate too much heat, melt the element, or make the transmission line sag down to an unsafe level.
	b) Voltage collapse: Voltages can become outside normal operating limits, and damage elements. 
	c) Other transient and dynamic oscillatory type behaviours that can affect the power system. 

For operating problems, operators monitor the constraints in real time, and identify cases when some constraints enter an insecure margin (Fig.~\ref{fig:interviews:why}a:~\textbf{\setlength{\fboxsep}{2pt}\colorbox{outlier}{Outlier} Identification}). 
To relieve the constraints, operators first propose possible solutions, summarise and compare the advantages and disadvantages of each, then choose the most suitable (Fig.~\ref{fig:interviews:why}b:~\textbf{\setlength{\fboxsep}{2pt}\colorbox{feature}{Feature} Summarisation and Comparison}).

For planning problems, the work flow is the same as demonstrated in Fig.~\ref{fig:interviews:why}a:~\textbf{\setlength{\fboxsep}{2pt}\colorbox{outlier}{Outlier} Identification}. Experts analyse the past records to identify where constraints occur and how frequently they occur.
They then modify the model to improve the compatibility for future years. This process usually combines analysing the trends of additional \emph{tabular} data, like demography (Fig.~\ref{fig:interviews:why}d:~\textbf{\setlength{\fboxsep}{2pt}\colorbox{trend}{Trends} Summarisation}). For extreme system collapse cases (e.g. tornado), analysis will be conducted afterwards with external \emph{tabular} data, like weather, to understand the reasons (Fig.~\ref{fig:interviews:why}a:~\textbf{\setlength{\fboxsep}{2pt}\colorbox{correlation}{Correlation} Identification}).

\vspace{-0.5em}
\subsection{Aircraft Trajectories}
\vspace{-0.2em}
\label{sec:interviews:airline} 
Aircraft trajectory data is used in both Air Traffic Control (ATC) and Management (ATM). ATC look at real-time aircraft trajectories to maintain traffic fluidity and security, while ATM analyses large amounts of past aircraft trajectories to understand past situations and improve future procedures~\cite{Cordeil:2016ed}. %
\deleted{removed: In July 2018, Maxime Cordeil was interviewed about his experience.} 

\noindent\textbf{What: Data Abstraction ---}
Information about each aircraft is collected during each flight, including: \textit{unique identifier}; \textit{3D Position}, i.e. longitude, latitude and altitude; \textit{timestamp}, \textit{speed, direction} and \textit{weather}, e.g. temperature, wind speed etc.
As sections of the airways have to avoid residential areas due to take off and landing noise, demography data is also usually used to inform the redesign of existing airways or design new ones.

\noindent\textbf{Why: Task Abstraction ---}
For ATC, the pilots and the air traffic controllers are actively communicating with each other on a regular basis. Some emergency cases were discussed in the interview:
\begin{itemize}
	\item Conflict between two aircrafts; i.e. if two aircraft retain their direction of flying, accidents can occur. Automatic systems are commonly used to warn of such a situation.
	\item Accidents happen on the runway; air traffic controllers will ask pilots to take an alternative runway for take off or landing.
	\item Weather; for landing in bad weather, pilots could be instructed to put on hold, and in some cases, to land at another nearby airport.
\end{itemize}
These actions involve ~\textbf{\setlength{\fboxsep}{2pt}\colorbox{outlier}{Outlier} Identification} (Fig.~\ref{fig:interviews:why}a) in both trajectories and other \emph{tabular} datasets, like weather. In some cases, other aircraft journeys may be affected, in which case urgent response proposals need to be derived. Air traffic controllers must summarise and compare different proposals to chose the most suitable (Fig.~\ref{fig:interviews:why}b:~\textbf{\setlength{\fboxsep}{2pt}\colorbox{feature}{Features} Summarisation and Comparison}).

In ATM, the main goal is to get a better understanding of unusual historical events and to avoid similar issues occurring in the future. Initally, aircraft or route trajectory outliers are identified. Then these outliers are correlated with other information, such as the aircraft condition or bad weather, to infer possible reasons (Fig.~\ref{fig:interviews:why}a:~\textbf{\setlength{\fboxsep}{2pt}\colorbox{correlation}{Correlation} Identification}). According to the interviewee, analysts mainly concentrate on individual trajectories, however they are also interested in the overview of a large number of trajectories, for example to investigate patterns in structure (Fig.~\ref{fig:interviews:why}c:~\textbf{\setlength{\fboxsep}{2pt}\colorbox{topology}{Topology} Identification}).

\vspace{-0.5em}
\section{Summary and Results}
\label{sec:interviews:summary}
\vspace{-0.2em}
In this section we aim to summarise the common analytical tasks identified in each of the five domains. From the interviews we identified six different \textbf{Consume} tasks and a number of \textbf{Produce} tasks. The \textbf{Produce} tasks usually involve additional modelling processes (e.g. \emph{deriving} constraints in energy network flow analysis). For completeness we described all tasks in the individual interviews in detail in Sec.~\ref{sec:interviews:details},
however, as our focus is on how to visually represent geographically-embedded flow data for analytical tasks, we concentrate on the \textbf{Consume} tasks for the rest of our analysis.  
The six \textbf{Consume} tasks are presented in Tab.~\ref{tab:interview:tasks}, revealling common tasks in each interview.  Notably, nearly all analysts were interested in \emph{Locate $\to$ Identify $\to$ Correlation or Outliers} tasks as well as \emph{Explore $\to$ Identify $\to$ Topology} tasks. Although this task taxonomy gives us a method to demonstrate the common themes between our interviews, it leds to over simplification of some of the tasks. When we look into the individual tasks we recognise that there is a difference in \emph{target} when performing tasks on geographically-embedded flow data. 

\begin{table}[t]
	\centering
    \includegraphics[width=\columnwidth]{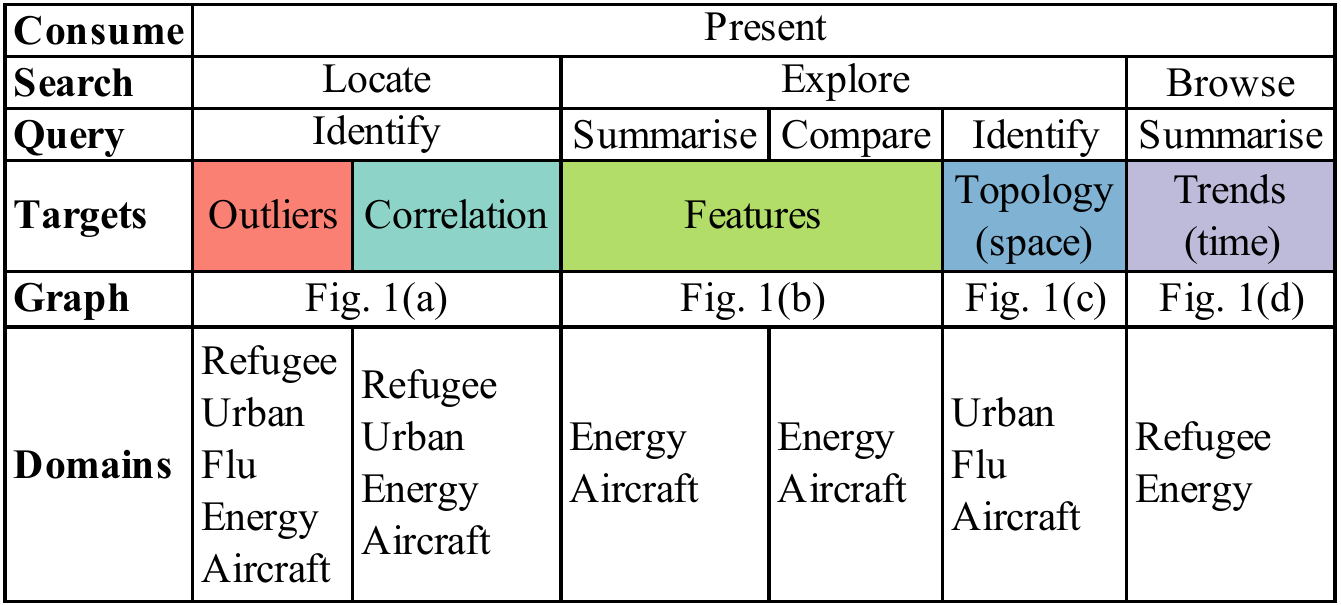}
	\caption{Tasks mentioned in each of the interviews.}
	\label{tab:interview:tasks}
	\vspace{-2em}
\end{table} 
 
Inspired by existing geovisualisation taxonomy research ~\cite{Andrienko:2006ek,Roth:2013ja}, we distinguish tasks into two levels: \emph{elementary} and \emph{synoptic}.
For the \emph{Locate $\to$ Identify $\to$ Correlation or Outliers} tasks, our domain experts are either interested in a flow between two locations (e.g. energy flow between two elements or the frequency of bicycle usage between two docking stations) or the total collection of flows into or out of a location (e.g. migrations into one country or number of bikes checked out from one docking station).
These can be considered as \emph{elementary tasks}.
In addition, the domain experts were also interested in the spatial patterns \emph{Explore $\to$ Identify $\to$ Topology} (e.g. the spatial movement of flu spread). This often involves the mental calculation of multiple flows into and out of multiple locations to see the regional pattern, or topology of the data. These can be considered as \emph{general tasks}.
We therefore identify that there are three \added{flow-targets} for analytical tasks relating to geographically-embedded flow data. These distinct differences in targets from the literature are fundamental in not only understanding the different tasks used in these domains, but also for describing the design space for geographically-embedded flow visualisation. To fill this gap, we propose three \added{flow-targets} for analysing \deleted{geographically-embedded} flow data:  
\begin{itemize}
	\item \emph{Single flow (SF)}: the magnitude of flow between two locations. 
\item \emph{Total flow (TF)}: the magnitude of flow in/out of a specific location. 
\item \emph{Regional flow (RF)}: the magnitude of flow between or within sets of locations. i.e. considering multiple locations as a single region. 
\end{itemize}
We categorise the tasks mentioned in interviews into these three \added{flow-targets} and found these targets were mentioned in almost all of the interviews (Tab.~\ref{tab:interview:targets}).
Notably the six different consume tasks (particularly \emph{features} and \emph{correlation}) appear in nearly all the three flow-targets. This helps demonstrates the importance of this multi-level target approach. Further evidence is shown through the specific reference to these different flow-targets in our evaluation of existing works and a new design for geographically embedded flow data~\cite{Yang:2017cy}.

\begin{table}[tb] 
\centering
    \includegraphics[width=\columnwidth]{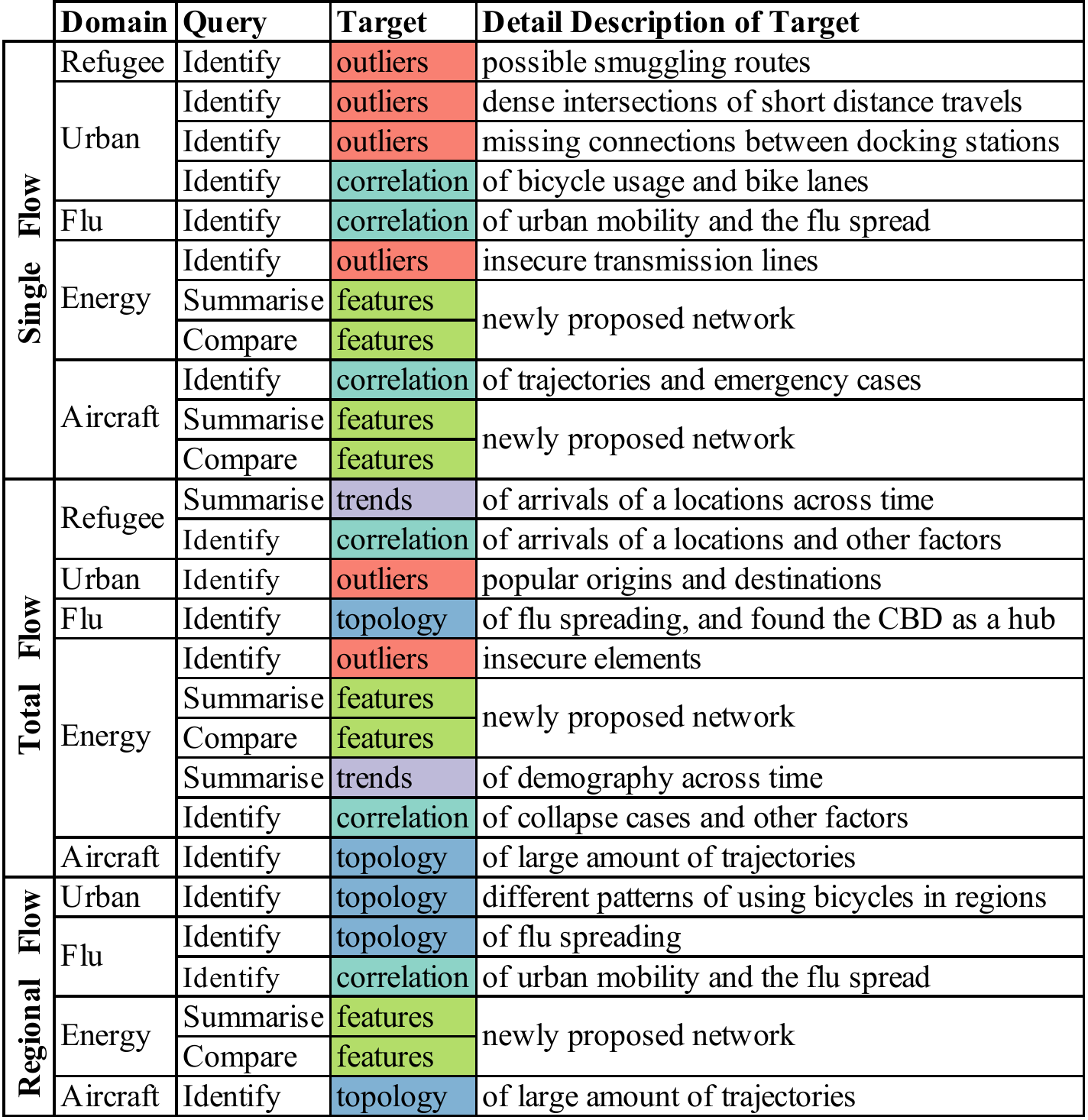}
	\caption{Tasks in the interviews categorised based on the targets of single flow (SF), total flow (TF) and regional flow (RF).
	}
	\label{tab:interview:targets}
	\vspace{-2em}
\end{table}

\vspace{-0.4em}
\section{Conclusion}
This paper analysed five expert interviews to better understand the real-world practice of analysing geographically-embedded flow data. Our analysis gives us a clearer understanding of the common tasks that analysts in different domains perform. Data and tasks discussed in the interviews were abstracted using Munzner’s what-why-how framework~\cite{Munzner:2014wj}. This was useful to structure our analysis of the interviews and demonstrate common tasks described by our interviewees, but we also recognised that a more detailed classification of the targets is required to better describe the role of flow map visualisation in the domains. Thus, we propose three flow-targets for analysing geographically-embedded flow data: \emph{single flow}, \emph{total flow} and \emph{regional flow}. We will continue to explore more domains and from more perspectives (e.g. scale and interactions).

\vspace{-0.5em}
\acknowledgments{
The authors wish to thank all interviewees for dedicating their time.
}

\bibliographystyle{abbrv-doi}

\bibliography{references}
\end{document}